\documentclass[amssymb,aps,pra,twocolumn,showpacs]{revtex4-1}
\usepackage{graphicx}

\begin{document}

\title{Spin-1 Atoms in Optical Superlattices: Single-Atom Tunneling
  and Entanglement}

\author{Andreas Wagner} 
\author{Christoph Bruder}
\affiliation{Department of Physics, University of Basel, 
Klingelbergstrasse 82, 4056 Basel, Switzerland}
\author{Eugene Demler}
\affiliation{Department of Physics, Harvard University, Cambridge,
  Massachusetts 02138, USA}

\date{December 20, 2011}

\begin{abstract}
We examine spinor Bose-Einstein condensates in optical superlattices
theoretically using a Bose-Hubbard Hamiltonian that takes spin
effects into account.  Assuming that a small number of spin-1 bosons
is loaded in an optical potential, we study single-particle tunneling
that occurs when one lattice site is ramped up relative to a
neighboring site. Spin-dependent effects modify the tunneling events
in a qualitative and quantitative way.  Depending on the asymmetry
of the double well different types of magnetic order occur, making the
system of spin-1 bosons in an optical superlattice a model for
mesoscopic magnetism. We use a double-well potential as a unit cell for
a one-dimensional superlattice.  Homogeneous and inhomogeneous
magnetic fields are applied and the effects of the linear and the
quadratic Zeeman shifts are examined.  We also investigate the bipartite
entanglement between the sites and construct states of maximal
entanglement. The entanglement in our system is due to both orbital
and spin degrees of freedom. We calculate the contribution of orbital
and spin entanglement and show that the sum of these two terms gives a
lower bound for the total entanglement.
\end{abstract}



\maketitle

\section{Introduction}
Ultracold atoms can be trapped via the ac Stark effect in optical
lattices, which are created by counterpropagating laser beams; in
case there are only a few atoms per site they build up optical
crystals. These systems offer a unique combination of experimental and
theoretical accessibilities~\cite{bloch08}. They can be manipulated with
a very high degree of accuracy and versatility so that they serve as
quantum simulators, i.e. they can be used to simulate complex problems
in many-body physics. Ultracold atoms in optical lattices offer robust
quantum coherence, a unique controllability and powerful read-out
tools, such as time-of-flight measurements~\cite{gerbier08,pedri01} or
fluorescence imaging~\cite{sherson10}.

Trapping ultracold atoms in conventional magnetic traps leads to
frozen spin degrees of freedom such that the atoms behave effectively
as spinless particles. If the atoms are trapped by optical means only,
the atoms keep the extra spin degree of freedom and the Bose-Einstein
condensate becomes a spinor condensate. The spinor degree of freedom
on alkaline gases corresponds to the manifold of degenerate Zeeman
hyperfine levels.  The ground-state properties of spinor Bose-Einstein
condensates in single traps were investigated in Refs.~\cite{ho98,ohmi98}.

We study the behavior of spin-1 atoms in optical superlattices,
in particular, an optical lattice that is formed by overlapping two
standing-wave laser fields with a commensurate wavelength ratio of 2.
The resulting lattice is an array of optical traps with a double-well
structure. We model the case when each double-well potential is filled
with a small number of spin-1 bosons. Spin-1 bosonic atoms in a 
double-well potential can be described by a variant of the two-site
Bose-Hubbard Hamiltonian~\cite{imambekov03}. This model allows examining the interplay between the kinetic energy (embodied by the
tunneling strength between the sites) and the particle interaction
(covered by the on-site interaction, i.e. the interaction within the
wells). Furthermore, it is possible to include an energy offset between
the sites, and the spin-1 Bose-Hubbard model additionally contains  a
term that incorporates spin-dependent interactions. This term
penalizes high-spin configurations on individual lattice sites in the case
of antiferromagnetic interaction between the atoms (e.g. for
$^{23}$Na) and low-spin configuration in the case of ferromagnetic
interactions (e.g. for $^{87}$Rb).

\begin{figure}[ht]
\begin{center}
\includegraphics[width=0.4\textwidth]{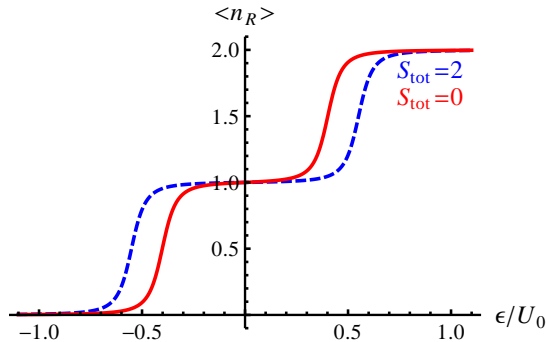}
\caption{(Color online) Two spin-1 bosons with antiferromagnetic ordering in a
  double-well potential. Here $n_R$ is the occupation number of the
  right well, and $\epsilon$ characterizes the energy offset between
  the two wells ($t/U_0=0.05$ and $U_2/U_0=0.1$).  Depending on the
  total spin of the system, bosonic staircase transitions occur at
  different bias voltages. Note that both states with $S_{tot}=0$  (red, solid line)
  and $S_{tot}=2$ (blue, dashed line) have symmetric orbital wavefunctions with 
  respect to particle exchange. The difference
  in the occupation numbers arises due to spin-dependent interactions
  and not due to a different orbital symmetry of the states. Thus, a
  measurement of the spin-dependent bosonic staircases provides a
  demonstration of mesoscopic magnetism.
\label{fig:2bosonsstaircase} }
\end{center}
\end{figure}

The two-site Bose-Hubbard model for spinless bosons can be used to
describe the transfer of single Cooper pairs in small Josephson
junctions, i.e. the physics of Cooper-pair
staircases~\cite{averin85,lafarge91,lafarge93}.  With ultracold
atoms in optical superlattices this model was realized and was
shown to give rise to a single-atom
staircase~\cite{gati07,averin08,cheinet08,ferrini08,rinck11}.  This
is achieved by monitoring the particle number in either of the wells
for different values of the energy offset.  In the case of small
tunneling strength, the difference in the number of atoms in the two
wells does not change smoothly when the energy offset is varied but
is characterized by a steplike behavior. Jumps from one step to
the next signal the tunneling of a single atom. In this paper, such
single-atom staircases are examined for spinor condensates.  Depending
on the energy bias, different types of magnetic order occur, and the
system of spin-1 bosons in an optical superlattice becomes a model for
mesoscopic magnetism.  A specific example of how this mesoscopic
magnetism can be observed in experiments is presented in
Fig.~\ref{fig:2bosonsstaircase}. This figure shows the difference
between bosonic staircases for two spin-1 bosons for configurations
with different total spins.  If the total spin is $S_{tot}=2$, the spins of
the two atoms are parallel and for antiferromagnetic interactions, as
in the case of $^{23}$Na, being in the same well costs extra
energy. Therefore the $S_{tot}=2$ configuration switches later (i.e. at
higher energy offset) to the state with both atoms in the same well.
In the ferromagnetic case (such as $^{87}$Rb), the curves for $S_{tot}=0$
and $S_{tot}=2$ will be exchanged.

Spin-1 atoms also allow stronger quantum correlations between the
wells compared with the case of spinless bosons. For spinless bosons
it has been noted that particle fluctuations between the left and the
right well lead to entanglement between the wells (see
e.g., Refs.~\cite{mazzarella,anna11} and references therein). In addition to this
orbital entanglement, spin-1 atoms allow spinor entanglement.  In this
paper, the quantum correlations between the wells are examined for
different values of the energy offset and different ratios of the
tunneling strength relative to the on-site interaction. We give a
lower bound for the entanglement between the wells by estimating the
amount of orbital and spinor entanglement separately.
At this point, we consider entanglement mainly as a theoretical
characterization of the many-body state of the system. 

The paper is organized as follows: In Sec.~\ref{sec:bhm}, the
two-site Bose-Hubbard model for spin-1 atoms is introduced and is given
explicitly for a small number of bosons. In Sec.~\ref{sec:staircases},
the physics of the bosonic staircases is discussed, and in Sec.~\ref{subsec:mag}, the effect of magnetic fields is included. In Sec.~\ref{sec:entanglement}, the bipartite entanglement for the two-site
Bose-Hubbard model is examined. The total entanglement between the
sites depends on orbital and spin degrees of freedom. We obtain a
lower bound of the total entanglement, which is given by the sum of
the orbital entanglement and the spin entanglement.

\section{Two-site Bose-Hubbard Hamiltonian for spin-1 atoms}
\label{sec:bhm}

The atoms we have in mind are alkali-metal atoms, such as $^{23}$Na and
$^{87}$Rb. Degenerate gases of alkali-metal atoms are weakly
interacting systems, but due to the confining lattice of
counterpropagating laser beams some of the atoms are forced to be
very close to each other and, thus, to become strongly interacting.  Spin-1
bosonic atoms in a double-well potential can be described by a
variant of the two-site Bose-Hubbard 
Hamiltonian~\cite{Jaksch1998,imambekov03},
\begin{eqnarray}
\label{mitspin}
H_0= \frac{U_0}{2} \sum_{i=L,R} n_i(n_i-1) -t 
\sum_{\sigma}(\hat L_{\sigma}^\dagger \hat R_{\sigma} 
+\hat R_{\sigma}^\dagger \hat L_{\sigma})  \nonumber \\ 
 +\varepsilon \left( n_L-n_R \right)+    
\frac{U_2}{2} \sum_{i=L,R}\left( \vec S_i^2 -2n_i\right),
\end{eqnarray}
where $\hat L_\sigma \ (\hat L_\sigma^\dagger)$ and $\hat R_\sigma
\ (\hat R_\sigma^\dagger)$ are annihilation (creation) operators for
atoms in the hyperfine state $\sigma \in \{ -1,0,1\}$ in the left or
right well and $n_L=\sum_\sigma L^\dagger_\sigma L_\sigma$
$\left(n_R=\sum_\sigma R^\dagger_\sigma R_\sigma \right)$ is the atom
number at the left (right) site.  The annihilation and creation
operators obey the canonical commutation relations
$[L_i,L^\dagger_j]=[R_i,R^\dagger_j]=\delta_{ij}$ and
$[R_i,L^\dagger_j]=[L_i,R^\dagger_j]=0$. $\vec S_L=\sum_{\sigma
  \sigma'} L^\dagger_\sigma \vec T_{\sigma \sigma'} L_{\sigma'}$ is
the total spin on the left site and the total spin on the right site
is $\vec S_R=\sum_{\sigma \sigma'} R^\dagger_\sigma \vec T_{\sigma
  \sigma'} R_{\sigma'}$, where $\vec T_{\sigma \sigma'}$ are the usual
spin-1 matrices.

The on-site repulsive interaction is described by the first term in 
Eq.~(\ref{mitspin}) and parametrized by $U_0$. 
The second term embodies the spin-symmetric
tunneling between the wells and $t$ is the hopping matrix
element between the lattice sites; $\varepsilon$ characterizes the
difference in on-site energy between the sites.  The term proportional
to $U_2$ describes spin-dependent interactions: 
It penalizes nonzero spin configurations
on individual lattice sites in the case of antiferromagnetic interactions
(e.g., $^{23}$Na) and favors high-spin configurations in the case of
ferromagnetic interactions (e.g., $^{87}$Rb).

The parameters can be controlled by adjusting the intensity of the
laser beams; it is possible to move from regimes of strong tunneling
($U_0 \ll t$) to regimes of very weak tunneling ($t \ll U_0$).  For
bulk lattices, it has been shown theoretically~\cite{Jaksch1998} and
experimentally~\cite{greiner02} that the system can be in a
Mott-insulating regime (for $t \ll U_0$) and in a superfluid phase,
where the kinetic energy dominates (for $U_0 \ll t$), and that it is
possible to switch from one regime to the other by tuning the laser
strength.

Whereas the ratio of $U_0/t$ can be changed, the ratio $U_0/U_2$ is
fixed for all lattice geometries. $U_2$ is given by the difference in
the scattering length of two spin-1 bosons in the case that their
spins couple to the total spin two, and the scattering length in the
case that their spins couple to the total spin zero.  This leads to an
estimated ratio of $U_2/U_0=0.04$ for $^{23}$Na~\cite{imambekov03}.

The Hamiltonian (\ref{mitspin}) conserves the particle number,
the $z$ projection of the total spin, i.e. $(\vec S_L + \vec S_R)_z$,
and the total spin $(\vec S_L + \vec S_R)^2$. In each well the bosonic
angular momenta couple, where symmetry constraints require $n_i + S_i$
to be even, and the resulting spins of both wells couple to a total angular
momentum $\vec S_{tot}=\vec S_L + \vec S_R$.

\subsection{Two spin-1 bosons}

Since the hopping term in Eq.~(\ref{mitspin}) conserves the absolute
value of the total spin $S_{tot}=|\vec S_L + \vec S_R|$, the Hilbert
space decomposes in orthogonal subspaces that do not mix, i.e.
\begin{eqnarray*}
\mathcal{H}=\mathcal{H}(S_{tot}=0)\oplus \mathcal{H}(S_{tot}=1) 
\oplus \mathcal{H}(S_{tot}=2)\:.
\end{eqnarray*} 
In the case of two spin-1 bosons, the Hilbert space is seven dimensional 
\begin{eqnarray*}
  |E_1 \rangle=|\{2,0\},\{0,0\},0\rangle,\\
  |E_2 \rangle=|\{1,1\},\{1,1\},0\rangle,\\
  |E_3 \rangle=|\{0,2\},\{0,0\},0\rangle,\\
  |E_4 \rangle=|\{1,1\},\{1,1\},1\rangle,\\
  |E_5 \rangle=|\{2,0\},\{2,0\},2\rangle,\\
  |E_6 \rangle=|\{1,1\},\{1,1\},2\rangle,\\
  |E_7 \rangle=|\{0,2\},\{0,2\},2\rangle,\\
\end{eqnarray*}
using the notation $|\{n_L,n_R\},\{S_L,S_R\},S_{tot}\rangle$. 
These basis vectors belong to three orthogonal subspaces,
\begin{eqnarray*}
\mathcal{H}=\underbrace{\{E_1,E_2,E_3\}}_{S_{tot} =0}\ \ 
\oplus \ \underbrace{\{E_4\}}_{S_{tot} =1} \ \ 
\oplus\ \ \underbrace{\{E_5,E_6,E_7\}}_{S_{tot} =2}\:.
\end{eqnarray*} 
To examine the ground-state properties of this system the Hamiltonian 
needs to be calculated and diagonalized for each subspace
separately. To calculate the off-diagonal elements of the Hamiltonian
it is necessary to write the elements of the whole system as
product of the single-well wavefunctions, e.g.,
\begin{eqnarray*}
| \{2,0\},\{2,0\},2\rangle &=& \\
| n_L=2,S_L=&2&, S_{1_z}=0\rangle  \otimes  | n_R=0,S_R=0,S_{2_z}=0\rangle
\end{eqnarray*}
and
\begin{eqnarray*}
|\{1,1\},&\{1,1\}&,2\rangle \\
&=& \sum_{m=-1,0,1}  C^{(2,0)}_{(1,m),(1,-m)} | 1,1,m\rangle 
\otimes | 1,1,-m\rangle \\
&=& \frac{1}{\sqrt{6}}(| 1,1,1\rangle \otimes | 1,1,-1\rangle \\
&+& 2 | 1,1,0\rangle\ \otimes | 1,1,0\rangle +| 1,1,-1\rangle 
\otimes | 1,1,1\rangle)
\end{eqnarray*}
where we have chosen the $S_z=0$ component for convenience because the
energy does not depend on the $S_z$ component. The single-well
wavefunctions need to be written in terms of single-particle creation
operators. For two spin-1 bosons this can be performed using the standard
Clebsch-Gordan coefficients. 

The diagonal elements are given by 
\begin{eqnarray*}
\langle E_1 |H |  E_1 \rangle &=&2 \epsilon +U_0-2 U_2, \\  
\langle E_3 |H |  E_3 \rangle &=&-2 \epsilon +U_0-2 U_2, \\ 
\langle E_5 |H |  E_5 \rangle &=&2 \epsilon +U_0+U_2,\\  
\langle E_7 |H |  E_7 \rangle &=&-2 \epsilon +U_0+U_2,\\
\langle E_2 |H |  E_2 \rangle &=&\langle E_4 |H |  E_4 \rangle
=\langle E_6 |H |  E_6 \rangle=0\:.  
\end{eqnarray*}
Due to the conservation of the total angular momentum the Hamiltonian
is block diagonal. The only nonvanishing tunneling elements are
\begin{eqnarray*}
\langle E_1 |H |  E_2 \rangle &=& \langle E_2 |H |  E_3 \rangle=-\sqrt{2} t, \\
\langle E_5 |H |  E_6 \rangle &=& \langle E_6 |H |  E_7 \rangle=-\sqrt{2} t.
\end{eqnarray*}

\subsection{Higher boson numbers}

The Hilbert space for three spin-1 bosons is given by the direct sum
of the following subspaces:
\begin{eqnarray*}
\mathcal{H}&=&\{E_1,E_2,E_3,E_4,E_{5},E_{6}\} \oplus \{E_7,E_8\} \\
&&\oplus \{E_9,E_{10},E_{11},E_{12}\}.
\end{eqnarray*}
The subspace $\{E_1,E_2,E_3,E_4,E_{5},E_{6}\}$ belongs to $|\vec S_L +
\vec S_R| =1$, the subspace $\{E_7,E_8\}$ belongs to $|\vec S_L + \vec
S_R| =2$, and the subspace $\{E_9,E_{10},E_{11},E_{12}\}$ belongs to
$|\vec S_L + \vec S_R| =3$.  The basis vectors are given by
 \begin{eqnarray*}
 |E_1\rangle &=|\{3,0\},\{1,0\},1\rangle, \\ 
 |E_2\rangle &=|\{2,1\},\{2,1\},1\rangle, \\ 
 |E_3\rangle &=|\{2,1\},\{0,1\},1\rangle, \\ 
 |E_4 \rangle &=|\{1,2\},\{1,2\},1\rangle,  \\ 
 |E_5 \rangle &=|\{1,2\},\{1,0\},1\rangle, \\ 
 |E_6 \rangle &=|\{0,3\},\{0,1\},1\rangle,\\
 |E_7 \rangle &=|\{2,1\},\{2,1\},2\rangle,  \\ 
 |E_8 \rangle &=|\{1,2\},\{1,2\},2\rangle, \\
  |E_9 \rangle &=|\{3,0\},\{3,0\},3\rangle,\\ 
  |E_{10} \rangle &=|\{2,1\},\{2,1\},3\rangle, \\ 
  |E_{11} \rangle &=|\{1,2\},\{1,2\},3\rangle, \\ 
  |E_{12} \rangle &=|\{0,3\},\{0,3\},3\rangle,
 \end{eqnarray*}
again using the notation $|\{n_L,n_R\},\{S_L,S_R\},S_{tot} \rangle$.

The Hamiltonian is block diagonal in the basis given above as in
the case of two bosons. The Hamiltonians belonging to $|\vec S_L
+ \vec S_R | =2$ and $|\vec S_L + \vec S_R | =3$ are quite similar to
the spinless case and the case of two spin-1 atoms. To calculate the
off-diagonal elements of the Hamiltonian we need to know how $n$
spin-1 bosons couple to a total spin $\vec S$ with a $z$-projection
$S_z$. For three spin-1 bosons this is performed in 
the Appendix.  The off-diagonal elements
for $|\vec S_L + \vec S_R | =2$ and $|\vec S_L + \vec S_R | =3$ are
\begin{eqnarray*}
\langle E_7 |H |  E_8 \rangle &=&- t, \\ 
\langle E_9 |H |  E_{10} \rangle &=& \langle E_{11} |H |  E_{12}
\rangle = -\sqrt{3} t,  \\ 
\langle E_{10} |H |  E_{11} \rangle &=& -2 t\:.
\end{eqnarray*}
The diagonal elements are given by
\begin{eqnarray*}
\langle E_7 |H |  E_7 \rangle &=&\langle E_{10} |H |  E_{10} 
\rangle=U_0+\epsilon +U_2, \\  
\langle E_{8} |H |  E_{8} \rangle &=&  \langle E_{11} |H |  E_{11} 
=U_0-\epsilon +U_2,\\
\langle E_9 |H |  E_9 \rangle &=&U_0+3 \epsilon +3 U_2, \\  
\langle E_{12} |H |  E_{12} \rangle &=& 3 U_0-3 \epsilon +3 U_2\:. 
\end{eqnarray*}
The Hamiltonian belonging to the Hilbert space $|\vec S_L + \vec
S_R | =1 $ exhibits a richer structure and differs from the spinless
case. This is because the term $ -t \sum_{\sigma}(L_{\sigma}^\dagger
R_{\sigma} + R_{\sigma}^\dagger L_{\sigma})$ describes tunneling
between several basis vectors, e.g. between $|E_1\rangle$ and
$|E_2\rangle$ as well as between $|E_1\rangle$ and $|E_3\rangle$. 
Because the energy does not depend on the
$S_z$ projection we can set $S_z=0$. The basis vector $|E_1\rangle$ is given
by
\begin{eqnarray*}
|\{3,0\},\{1,0\},1\rangle &=& |3,1,0\rangle \otimes |0,0,0\rangle \\
&=&\left[ \sqrt{\frac{2}{5}}  \hat L_{-1}^\dagger  \hat L_0{}^\dagger  \hat L_1^\dagger-\sqrt{\frac{1}{10}} \left( \hat L_0^\dagger\right)^3 \right]  |0\rangle,
\end{eqnarray*}
and the basis vector $|E_2\rangle$ is given by
\begin{eqnarray*}
&|\{2,1\}&,\{2,1\},1\rangle   \\
&=&\sum_{m=-2,\ldots,2}  \sum_{n=-1,0,1} C^{(1,0)}_{(1,m),(1,n)} | 2,2,m\rangle \otimes | 1,1,n\rangle \\
&=&\sqrt{\frac{3}{10}} |2,1,-1\rangle\otimes|1,1,1\rangle - \sqrt{\frac{4}{10}} |2,1,0\rangle\otimes|1,1,0\rangle \\
&+&\sqrt{\frac{3}{10}}  |2,1,1\rangle\otimes|1,1,-1\rangle \\
&=& \Bigg[ -\sqrt{ \frac{2}{15} } \left( \hat L_0^\dagger \right)^2  \hat R_0^\dagger +\sqrt{\frac{3}{10}} \hat L_1^\dagger \hat L_0^\dagger \hat R_{-1}^\dagger \\
&+& \sqrt{\frac{3}{10}} \hat L_{-1}^\dagger \hat L_0^\dagger \hat R_1^\dagger-\sqrt{\frac{2}{15}} \hat L_{-1}^\dagger \hat L_1^\dagger \hat R_0^\dagger \Bigg]  |0\rangle.
\end{eqnarray*}
Now, we can calculate the corresponding nondiagonal element of the Hamiltonian,
\begin{eqnarray*}
\langle E_1 | H | E_2 \rangle=-t \ \langle E_1| 
\sum_{\sigma}(\hat L_{\sigma}^\dagger \hat R_{\sigma} + 
\hat R_{\sigma}^\dagger \hat L_{\sigma})|E_2 \rangle=-\sqrt{\frac 5 3} \ t.
\end{eqnarray*}
The basis vector $|E_3\rangle$ is given by
\begin{eqnarray*}
|\{2,1\},\{0,1\},1\rangle&=& |2,0,0\rangle\otimes|1,1,0\rangle\\
&=& \left[ \sqrt{\frac{2}{3}}\hat L_{-1}^\dagger \hat L_1^\dagger 
\hat R_0^\dagger-\frac{1}{\sqrt{6}} \left( \hat L_0^\dagger\right)^2 
\hat R_0^\dagger\right]  |0\rangle\:, 
\end{eqnarray*}
and the corresponding nondiagonal element of the Hamiltonian is 
\begin{eqnarray*}
\langle E_1 | H | E_1 \rangle = -\sqrt{\frac 4 3} \ t.
\end{eqnarray*}
Note that the off-diagonal elements of the Hamiltonian depend on the
spin configurations, also, the off-diagonal elements do not depend on
the strength of the spin-dependent interactions $U_2$.  Similar
calculations lead to the remaining off-diagonal elements,
\begin{eqnarray*}
\langle E_2 | H | E_4 \rangle &=&-\frac{t}{3}, \\ 
\langle E_2 | H | E_5 \rangle &=&\langle E_3 | H | E_4 
\rangle=-\frac{2 \sqrt{5}t}{3}, \\ 
\langle E_3 | H | E_5 \rangle &=&-\frac{2 t}{3},\\
\langle E_4 | H | E_6 \rangle &=&\langle E_1 | H | E_2 
\rangle=-\sqrt{\frac{4}{3}}, \\ 
\langle E_5 | H | E_6 \rangle &=&\langle E_1 | H | E_3 
\rangle=- \sqrt{\frac{5}{3}}\:.
\end{eqnarray*}
The diagonal elements are given by 
\begin{eqnarray*}
\langle E_1 |H |  E_1 \rangle &=3 U_0+3 \epsilon -2 U_2, \\  
\langle E_{2} |H |  E_{2} \rangle &= U_0+\epsilon +U_2,\\
\langle E_3 |H |  E_3 \rangle &=U_0+\epsilon -2 U_2, \\  
\langle E_{4} |H |  E_{4} \rangle &= U_0-\epsilon +U_2, \\
  \langle E_{5} |H |  E_{5} \rangle &=U_0 -\epsilon -2 U_2, \\ 
\langle E_{6} |H |  E_{6} \rangle &= 3 U_0-3 \epsilon -2 U_2\:. 
\end{eqnarray*}

Higher boson numbers lead to analogous expressions that are used in
the following but are not given here.

\section{Bosonic staircases}
\label{sec:staircases}

The two-site Bose-Hubbard model may be used to model Cooper-pair
staircases~\cite{averin85,lafarge91,lafarge93} relevant for small
Josephson junctions. In the case of ultracold atoms the same effect
gives rise to single-atom
staircases~\cite{gati07,averin08,cheinet08,ferrini08,rinck11}. Here
we present such staircases for spin-1 atoms.

\subsection{General treatment}
\label{subsec:general}

The Hilbert spaces decompose into different subspaces according to the total
spin of the system. Different subspaces are not mixed by ramping up the
energy difference between the double wells and behave in a
different way. In the case of two bosons this is shown in
Fig.~\ref{fig:2bosonsstaircase}. The different widths of the steps
centered at $\epsilon =0$ correspond to different values of total
spins per site. At $\epsilon =0.5$, state
$|\{1,1\},\{1,1\},2\rangle$ is energetically lower than state
$|\{0,2\},\{0,2\},2\rangle$, which makes the step broader for $S_{tot}=2$. 
On the contrary, at $\epsilon =0.5$, state $|\{1,1\},\{1,1\},0\rangle$ is
energetically higher than state $|\{0,2\},\{0,0\},0\rangle$, which
makes the step narrower for $S_{tot}=0$.

In general, depending on the sign of $U_2$, states with high
single-well angular momenta get penalized or get favored. If $U_2 > 0$
(such as, e.g., for $^{23}$Na), nonzero spin configurations get penalized. In the
case of $^{87}$Rb, $U_2$ is negative and spin-dependent interactions
lead to the opposite effect: High-spin configurations are favored, and
the corresponding steps are broader.
Therefore, in the ferromagnetic case, the curves for $S_{tot}=0$ and
$S_{tot}$ in Fig.~\ref{fig:2bosonsstaircase} will be exchanged.

The exact position of the steps can be calculated in the atomic
limit, i.e., $t=0$. The step positions generally depend  linearly on
$U_2$. For some spin configurations, e.g., odd atom number, lowest
possible total spin, and antiferromagnetic interactions, the step
positions do not depend on spin-dependent interactions.

\begin{figure}[h]
\begin{center}
\includegraphics[width=0.4\textwidth]{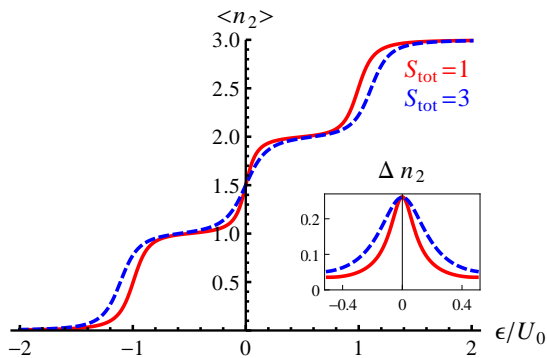}
\caption{(Color online) Bosonic staircase for three spin-1 bosons with
  antiferromagnetic ordering in a double-well potential 
($t/U_0=0.05$ and $U_2/U_0=0.1$):   $S_{tot}=1$ (red, solid line) and  
$S_{tot}=3$ (blue, dashed line). (Inset) Variance in the particle number
  in the left well for the step around
  $\epsilon=0$.\label{Fig:3bosonsstaircase} }
\end{center}
\end{figure}

For higher boson numbers, the richer structure of the off-diagonal
elements means that the variance in the particle number
depends on the total spin and the energy offset. In the case of three
bosons (Fig.~\ref{Fig:3bosonsstaircase}), the step at $\epsilon=0$ is
not shifted due to symmetry reasons, whereas the steps at $\epsilon=1$
and $\epsilon=-1$ are shifted linearly. At the same time, the steps
belonging to $S_{tot}=3$ are not as sharp as the steps belonging to
$S_{tot}=1$, i.e., the curve of the variance in $n_L$ is broader in the
case of $S_{tot}=3$.

\begin{figure}[h]
\includegraphics[width=0.4\textwidth]{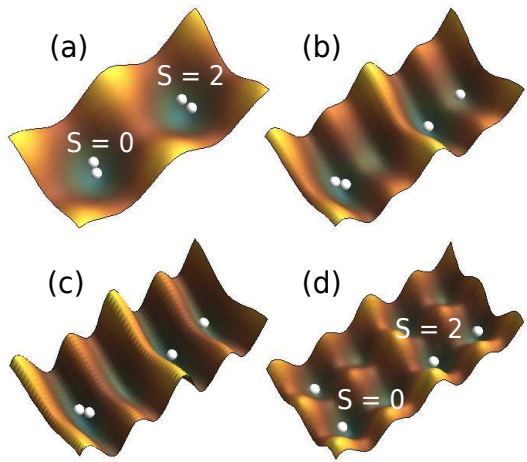}
\caption{(Color online) A possible way to separate the $S_{tot}=2$ spin component from the $S_{tot}=0$ spin component in the case of 
  antiferromagnetic interactions between the atoms: The potentials 
  in the $x$ and $y$ directions are manipulated separately. In the first
  step (b), the
  energy offset between the double wells is lifted until the bosons
  combining with the total spin $S_{tot}=0$ separate while the bosons
  belonging to $S_{tot}=2$ still remain in the same site. (c) Next, the
  wells are separated by a large potential barrier and tunneling is
  suppressed. (d) An additional laser is switched on and the
  bosons coupling to $S_{tot}=2$ distribute in the resulting double
  well. The switching is assumed to happen adiabatically such that the 
  system can be regarded to be in the ground state at every instant. 
\label{fig:two}
}
 \end{figure}

The staircases for different total spins may be used to arrange spin-1
atoms in a two-dimensional (2D) superlattice according to their spin
degrees of freedom (see Fig.~\ref{fig:two}).

\subsection{Beyond ground-state analysis}

\begin{figure}[h]
\begin{center}
\includegraphics[width=0.4\textwidth]{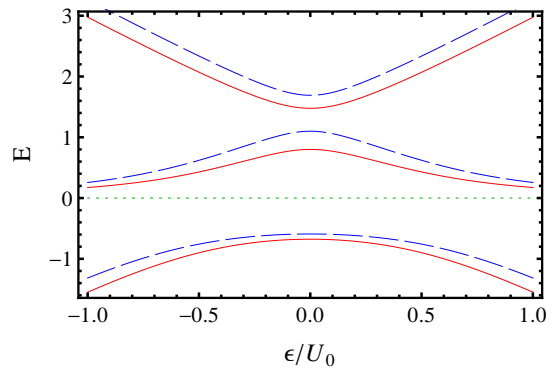}
\caption{(Color online) Energy spectrum of two spin-1 bosons in a
  double well with strong tunneling ($t/U_0=0.5$ and $U_2/U_0=0.1$):
  $S_{tot}=0$ subspace (red, solid lines), $S_{tot}=2$ (blue, dashed
  lines), and $S_{tot}=1$ (dotted line). \label{fig:spectrum1} }
\end{center}
\end{figure}

\begin{figure}[h]
\begin{center}
\includegraphics[width=0.4\textwidth]{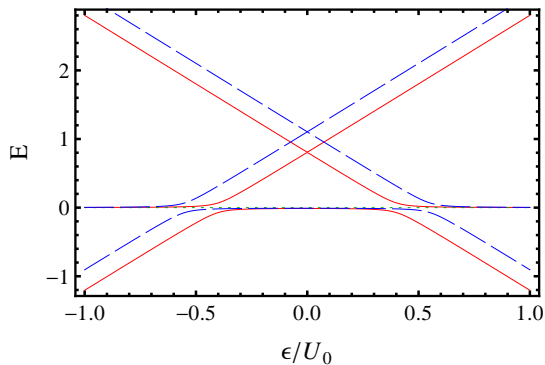}
\caption{(Color online) Energy spectrum of two spin-1 bosons in a
  double well with weak tunneling ($t/U_0=0.05$ and $U_2/U_0=0.1$).
  Color code as in Fig.~\ref{fig:spectrum1}.
\label{fig:spectrum2} }
\end{center}
\end{figure}

The gap between the ground state and the first excited state in the
energy spectrum depends strongly on the tunneling between the sites
(see Figs.~\ref{fig:spectrum1} and \ref{fig:spectrum2}). For
finite temperatures the density matrix describing the system, thus, is highly mixed for small tunneling parameters, and the ground-state
behavior only dominates if tunneling is sufficiently strong.

\subsection{Magnetic field included}
\label{subsec:mag}
\begin{figure}[th]
\includegraphics[width=0.4\textwidth]{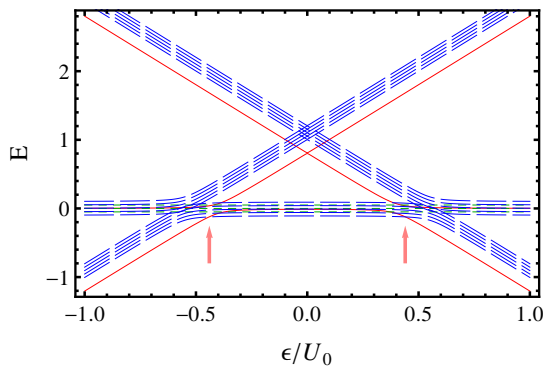}
\caption{(Color online) Linear Zeeman shift of the energy levels in the energy
  spectrum of two spin-1 bosons ($B/U_0 = 0.05$, $t/U_0=0.05$ and
  $U_2/U_0=0.1$).  The energy levels of Fig.~\ref{fig:spectrum2} split
  into spin multiplets. The red arrows denote ground state level
  crossings. Color code as in Fig.~\ref{fig:spectrum1}.
\label{fig:spectrum-mag} }
\end{figure}

\begin{figure}[h]
\begin{center}
\includegraphics[width=0.4\textwidth]{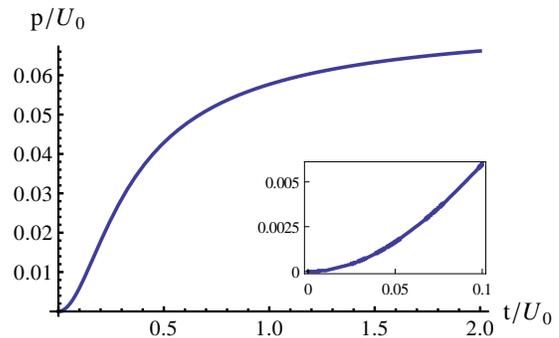}
\caption{(Color online) Critical magnetic field $p=g \mu_B B$ above which the
  staircase for two bosons shows a discontinuous behavior signifying
  spin-flip transitions ($U_2/U_0=0.1$). \label{fig:magcritical} }
\end{center}
\end{figure}

The effect of a magnetic field can be included in the model
(\ref{mitspin}) by adding a term to the Hamiltonian thath describes
the coupling of the spins to the magnetic
field~\cite{imambekov04}. The first contribution of a magnetic field
$\vec B= (0,0,B)$ is a regular Zeeman shift in the energy levels:
\begin{eqnarray*}
H = H_0 + p \sum_{i=L,R}  \sum_{\sigma}m_{i \sigma} \hat n_{i \sigma} =   H_0 + p \ S^{tot}_z
\end{eqnarray*}
where $p=g \mu_B B$ and $ \hat n_{i \sigma}$ is the particle
number operator for the $i$th site that gives the number of bosons in
the $m$th hyperfine state. The linear Zeeman shift changes the overall
state considerably. The energy eigenvalues belonging to $S_{tot} \neq
0$ split into multiplets because the hyperfine levels are
no longer degenerate (see Fig.~\ref{fig:spectrum-mag}).
For a given tunneling strength, there is a critical magnetic-field
strength that leads to ground-state level crossings. Such level
crossings correspond to spin-flip transitions, i.e., the
ground-state energy is continuous, but the expectation values of the
particle number of single sites and of the magnetization are not. 
This means that the
overall ground state of the system does not belong to the same
$z$ projection of the total spin for all values of the energy offset
$\epsilon$. Figure~\ref{fig:magcritical} shows the critical value of the
magnetic field in the case of two bosons.

However, spin-non-conserving collisions are negligible over the
lifetime of the condensate, and the total magnetization is a conserved
quantity on the time scale of the experiment~\cite{stenger98,
  rodriguez11}.  For a given magnetization the properties of the
system are not altered by the linear Zeeman effect; the whole spectrum
merely is shifted. Only if one is interested in comparing different
magnetizations, the linear Zeeman effect has to be taken into account.
In a series of experiments with a given magnetization, it is necessary,  therefore, to include higher-order contributions in the magnetic field.
The quadratic Zeeman effect arises because the hyperfine spins characterizing
ultracold atoms are mixtures of electron and nuclear spins. Since the
magnetic field approximately  couples only to the electron spin, the
Zeeman effect is nonlinear in the field but ,typically, can be described
by a sum of linear and quadratic terms.

For each of the subspaces belonging to different
magnetizations $S^{tot}_z$, there is a separate effective Hamiltonian
\begin{eqnarray}
H_q &=& H_0 + q \sum_{i=L,R}  \sum_{\sigma}m_{i \sigma}^2 \hat n_{i \sigma}\:.
\label{mitspinandmag}
\end{eqnarray}
The magnitude of the quadratic Zeeman shift is given by 
$q=q_0 B^2$, where e.g. $q_0=h \times 390 $ Hz/G$^2$ for 
Na~\cite{stenger98}. 

\begin{figure}[ht]
\begin{center}
\includegraphics[width=0.4\textwidth]{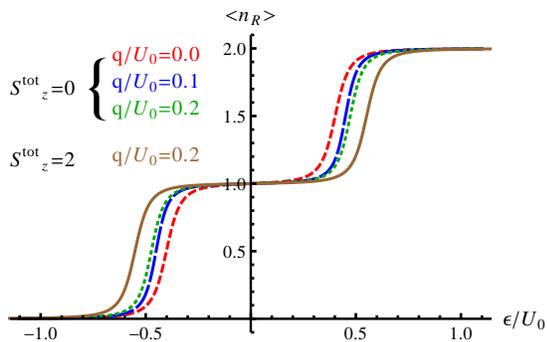}
\caption{(Color online) Two spin-1 bosons with antiferromagnetic
  ordering in a double-well potential ($t/U_0=0.05$ and
  $U_2/U_0=0.1$).  (Dashed lines) $S^{tot}_z=0$ for different magnetic
  fields $q=q_0 B^2$ (short dashes q/$U_0$=0.2, long dashes
  q/$U_0$=0.1 and medium sized dashes q/$U_0$=0).  (Solid line)
  $S^{tot}_z=2$. In this case the staircase does not depend on the
  magnetic field. The difference in this staircase to the ones with
  $S^{tot}_z=0$, which is the main manifestation of mesoscopic
  magnetism, persists in the presence of the quadratic Zeeman effect.
\label{fig:2bosonsstaircase-mag} }
\end{center}
\end{figure}

In the case of two bosons the system with magnetization $S^{tot}_z=0$
possesses the most interesting structure because the Hilbert space is
composed of states with different total spins.  For $S^{tot}_z=2$, the
quadratic Zeeman shift does not alter the staircase since it
leads to a homogeneous shift in all the energy levels.  The 
staircases at different magnetic fields are shown in
Fig.~\ref{fig:2bosonsstaircase-mag}.  For $S^{tot}_z=0$, the step
positions depend in a nonlinear way on the magnetic field strength.
It is no longer possible to read them off in the atomic limit
(i.e., $t=0$) because the existence of the quadratic Zeeman shift
leads to additional nondiagonal elements in the Hamiltonian.  Note
that the quadratic Zeeman effect does not eliminate the difference in
the two staircases, which is the main manifestation of mesoscopic
magnetism.

\begin{figure}[h]
\includegraphics[width=0.49\textwidth]{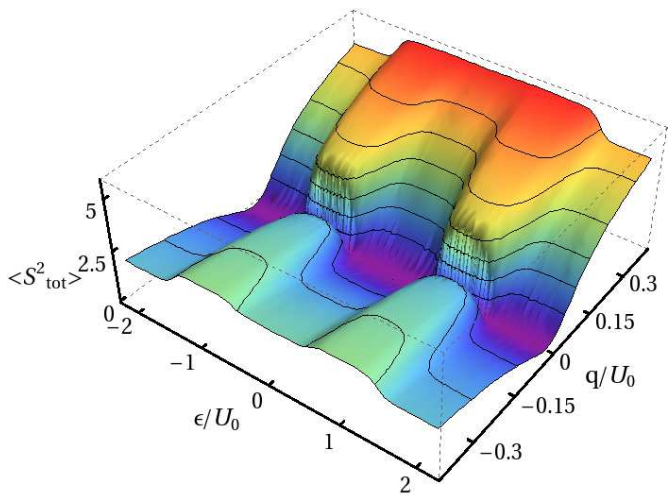}
\caption{(Color online) Expectation value $\langle S_{tot}^2 \rangle$ of the system
  for four bosons for different magnetic fields $q=q_0 B^2$ 
($S^{tot}_z=0$, $t/U_0=0.05$ and $U_2/U_0=0.1$).  
\label{fig:stot-mag-3d} }
\end{figure}

Due to the fact that the quadratic Zeeman shift does not commute with
the operator of the total spin $S^{tot}$, the eigenstates of the
Hamiltonian given in Eq.~(\ref{mitspinandmag}) are no longer
eigenstates of $S_{tot}$. For $B \neq 0$, the ground state of the
system is a superposition of different eigenstates of $S_{tot}$,
i.e., states with different $S_{tot}$ hybridize (see
Fig.~\ref{fig:stot-mag-3d}).  For certain values of the energy offset
$\epsilon$ (e.g., $\epsilon/U_0=1$ and $\epsilon/U_0=-1$ for four
bosons), the appearance of a magnetic field changes the ground state
strongly. This reflects the specific spin configurations.

The quadratic Zeeman shift also changes the overall spectrum for a
given magnetization qualitatively such that, in the case of thermal
occupation of the double well the density matrix of the system changes
considerably. For $q=0$, the ground state nearly is degenerate with the
first excited state, whereas the gap widens for finite values of $q$.

Additionally one can include inhomogeneous magnetic fields,
\begin{eqnarray*}
H &=& H_q + \Delta B  \left(  S_{Lz} - \  S_{Rz} \right)\:, 
\end{eqnarray*}
where $\Delta B$ describes the strength of the field gradient. The
magnetic-field offset $\Delta B$ changes the Hamiltonian if $S_{tot_z}
\neq 0$. For some configurations, e.g.  two bosons in a double
well, $\Delta B$ merely leads to an overall shift in $\epsilon$,
i.e., an inhomogeneous magnetic field is equivalent to an energy
offset $\epsilon$. In general, this is not the case, and $\Delta B$  is an 
additional tool to reshape the staircases depending on the
spin configuration of the system.

\section{Entanglement for spin-1 bosons}
\label{sec:entanglement}

Entanglement is a unique feature of quantum-mechanical
systems. Understanding entanglement deepens our understanding of
quantum mechanics and, therefore, is of fundamental interest. Moreover,
entanglement is a resource for quantum computation and correlates
separated systems stronger than all classical correlations can do.
For bipartite pure states entanglement is well understood and the
different entanglement measures are equivalent. In the following we
will use the entanglement of formation (EOF)~\cite{popescu97} as an
entanglement measure. The EOF is the number of Einstein-Podolsky-Rosen
pairs asymptotically required to prepare a given state by local
operations and classical communication. The entanglement of formation
between two qudits ($D$-dimensional objects) in a pure state is given
by the von Neumann entropy of the reduced density matrix of each
single qudit~\cite{dennison01}.  To calculate the EOF, consider a
system consisting of two parts labeled $A$ and $B$. Any pure state
$|\Psi \rangle$ of the system can be written in the Schmidt
decomposition~\cite{horodecki09}
\begin{eqnarray*}
|\Psi \rangle = \sum_i^D  c_i |\psi_i^A \rangle \otimes |\psi_i^B \rangle\:,
 \end{eqnarray*}
where $\{\psi_1^A, \ldots, \psi_D^A \}$ 
and $\{\psi_1^B, \ldots, \psi_D^B \}$ are complete sets
of orthonormal states of the respective subsystems. The system is entangled iff there is more than one nonvanishing coefficient $c_i$. These coefficients $c_i$ are positive, unique and invariant under local operations and, therefore, can  be used to
quantify the entanglement between $A$ and $B$. The von Neumann entropy
of the reduced density matrix of each single qudit is given by
\begin{eqnarray*}
E(\Psi)&=&S(\text{Tr}_B |\Psi \rangle \langle \Psi |)=
S(\text{Tr}_A |\Psi \rangle \langle \Psi |)\\
&=&-\sum_1^D c_i^2 \log_2 c_i^2,
\end{eqnarray*}
where $S$ indicates the entropy. It ranges from zero to $\log_2
D$. The entanglement of formation of two qudits with $D>2$ thereby
exceeds the entanglement of formation of two qubits, i.e. 
higher-dimensional objects contain more entanglement and violate all
Clauser-Horne-Shimony-Holt-inequalities more strongly than qubits.

In the following we calculate the EOF in a double well and
examine how much the two sites are entangled.
At this point, we consider the EOF mainly as a theoretical
characterization of the many-body state of the system. 

\begin{figure}[ht]
 \begin{center}
 \includegraphics[width=0.4\textwidth]{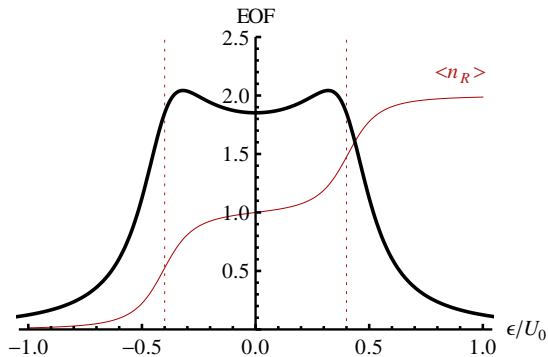}
 \caption{(Color online) EOF between two wells for two
   bosons with antiferromagnetic interactions ($t/U_0=0.1$ and
   $U_2/U_0=0.1$) for the total spin $S_{tot}=0$. \label{fig:ent2b1} }
 \end{center}
 \end{figure}

The EOF for typical parameters is shown in
Fig.~\ref{fig:ent2b1}. 
The maximal entanglement exceeds the maximal entanglement between two
qutrits of $\log_2 3\approx 1.585$. This is due to particle fluctuations. The
total amount of entanglement stems from orbital and spin degrees of
freedom. 

\begin{figure}[h]
 \begin{center}
 \includegraphics[width=0.5\textwidth]{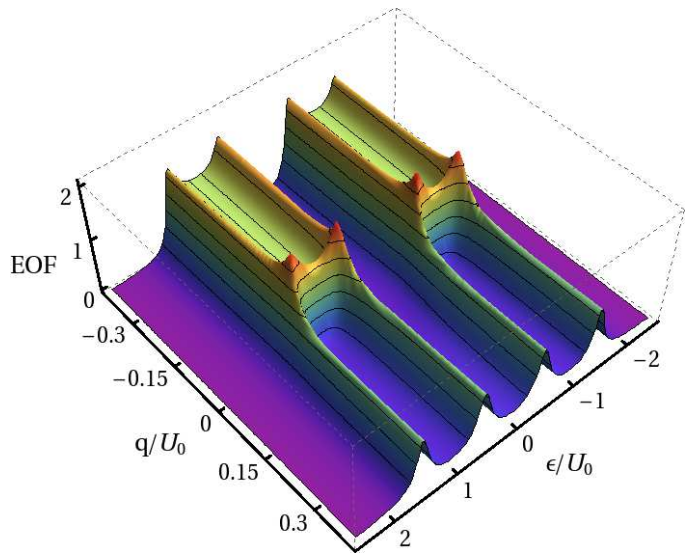}
\caption{(Color online) EOF between two wells for four particles in the presence of a 
magnetic field ($S_z^{tot}=0$, $t/U_0=0.05$ and $U_2/U_0=0.1$). 
For $q>0$, even small fields will eliminate the contribution of 
the spin degrees of freedom to the entanglement.
\label{fig:ent4b3d} }
 \end{center}
 \end{figure}

Magnetic fields have a strong effect on the entanglement of
formation. Figure~\ref{fig:ent4b3d} shows the EOF of four bosons at
$S_z^{tot}=0$.  For $q>0$, the contribution of the spin degrees of
freedom to the entanglement of formation is suppressed already by
small magnetic fields.  This is somewhat surprising, because the
system is constrained to $S^{tot}_z$, i.e. the state with the
strongest spin entanglement of all states with a given total spin.
For $q<0$, this contribution initially is  reduced but then remains
constant as a function of $q$.

\subsection{Two spin-1 bosons}
For two bosons and in the case of $S_{tot}=0$ a possible orthonormal
basis is given by
\begin{eqnarray*}
&\{ \psi_1&,\psi_2,\psi_3\}\\
&=&\{ |\{2,0\},\{0,0\},0\rangle, |\{1,1\},\{1,1\},0\rangle,|\{0,2\},\{0,0\},0\rangle \}
 \end{eqnarray*}
using the notation $|\{n_L,n_R\},\{S_L,S_R\},S_{tot} \rangle$. The
decomposition,
\begin{eqnarray*}
|\Psi \rangle = \sum_i^3  c_i |\psi_i \rangle
 \end{eqnarray*}
is not a Schmidt decomposition, because the vector $| \psi_2 \rangle$
is a superposition of orthonormal states,
\begin{eqnarray*}
|\{1,1\},\{1,1\},0\rangle&=&-\frac{1}{\sqrt{3}}|1,1,0;1,1,0\rangle \\
+ \frac{1}{\sqrt{3}}|&1,&1,1;1,1,-1\rangle + 
\frac{1}{\sqrt{3}}|1,1,-1;1,1,1\rangle,
\end{eqnarray*}
using the notation $|n_L,S_L,S_{Lz};n_R,S_R,S_{Rz}\rangle$. The
entanglement of formation of $|\psi_2\rangle$ is given by
$$ E(|\psi_2\rangle)=3 \frac 1 3 \log_2 3\:.$$
The EOF of $|\Psi \rangle$ is given by
\begin{eqnarray}
\label{2bedecompose}
&E&(|\Psi \rangle)  \nonumber  \\
&=& - c_1^2 \log_2 c_1^2 - c_3^2 \log_2 c_3^2 - 3 \left( c_2 \frac{1}{\sqrt{3}} \right)^2 \log_2 \left( c_2 \frac{1}{\sqrt{3}} \right)^2  \nonumber  \\
&=&- \sum_i^3  c_i^2 \log_2 c_i^2 + c_2^2 \log_2 3    \nonumber \\
&=&- \sum_i^3  c_i^2 \log_2 c_i^2 + \sum_i^3  c_i^2 E(|\psi_i\rangle)   \nonumber \\
&=& E_{\text{orbital}} + E_{\text{spin}}.
\end{eqnarray}
The total entanglement between the left and the right well decomposes
in an orbital part and a spin part. The orbital part stems from the
coefficients that distinguish different orbital wave functions. The
spin part originates from the EOF of the individual
basis vectors, each weighted with the coefficient $c_i^2$. The
coefficients $c_i$ depend on the tunneling strength $t$, on the
on-site interaction $U_0$, on the spin-dependent interaction $U_2$, and on
the energy offset $\varepsilon$. 

\begin{figure}[ht]
\includegraphics[width=0.4\textwidth]{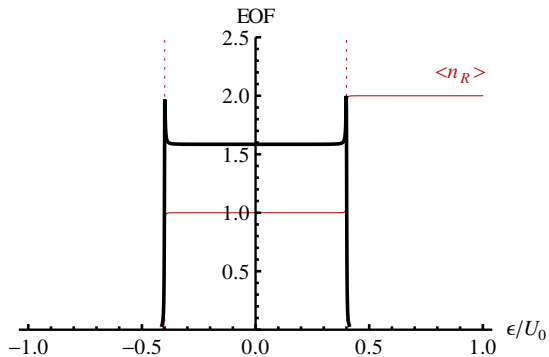}
\caption{(Color online) EOF between two wells for two bosons with
  very weak tunneling and antiferromagnetic interactions 
($t/U_0=0.001$ and $U_2/U_0=0.1$) for the total spin
  $S_{tot}=0$.\label{fig:ent2b2} }
 \end{figure}

In the limit of weak tunneling $t \ll U_0$ the Hamiltonian is diagonal
in the basis $\{ \psi_1,\psi_2,\psi_3\}$ and the ground state of a
symmetric double-well potential (i.e. $\varepsilon = 0$) is given by
\begin{eqnarray}|
\Psi_0\rangle &=& |\psi_2\rangle=-
\frac{1}{\sqrt{3}}|1,1,0;1,1,0\rangle   \nonumber \\
+& \frac{1}{\sqrt{3}}&|1,1,1;1,1,-1\rangle + 
\frac{1}{\sqrt{3}}|1,1,-1;1,1,1\rangle\:,
\end{eqnarray}
which leads to an entanglement of $E(|\psi_2\rangle)=\log_2 3$, see
Fig.~\ref{fig:ent2b2}. 

\begin{figure}[h]
 \begin{center}
 \includegraphics[width=0.4\textwidth]{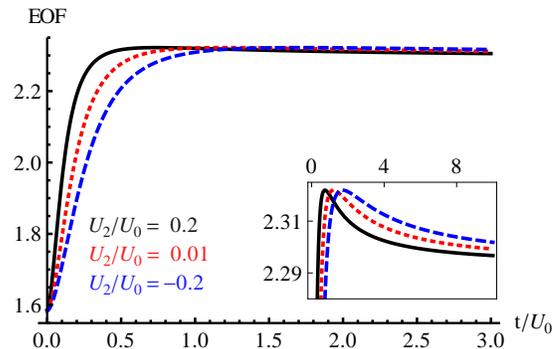}
\caption{(Color online) EOF between two symmetric wells ($\varepsilon
  =0$) for different values of $U_2$, i.e., different spin
  interactions (solid line $U_2/U_0=0.2$, dotted line $U_2/U_0=0.01$,
  and dashed line $U_2/U_0=-0.2$).\label{fig:ent2b3} }
 \end{center}
 \end{figure}

\begin{figure}[h]
 \begin{center}
\includegraphics[width=0.4\textwidth]{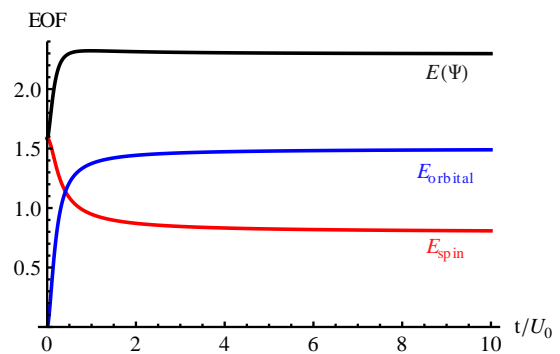}
\caption{(Color online) EOF, $E_{\text{spin}}$ and
  $E_{\text{orbital}}$ between two symmetric wells for $U_2/U_0 =
  0.1$.\label{fig:ent2b5} }
 \end{center}
 \end{figure}

In the limit of strong tunneling (i.e. $U_0 \ll
t$), the ground state of the system is
\begin{eqnarray}
|\Psi_0\rangle=\frac{1}{2}|\psi_1\rangle+
\frac{1}{\sqrt{2}}|\psi_2\rangle+\frac{1}{2}|\psi_3\rangle.
\end{eqnarray}
For this state the orbital entanglement is given by $E_{\text{orbital}}=-2
\frac 1 4 \log_2 \frac 1 4 - \frac 1 2 \log_2 \frac 1 2=3/2$ 
and the spin entanglement is given by
$E_{\text{spin}}=\left(\frac{1}{\sqrt{2}}\right)^2  \log_2
3\approx 0.792$. Therefore, the total entanglement is 
$E(|\Psi_2\rangle)\approx 2.292$. This is not the maximum amount of
entanglement that can be obtained for this system (see
Fig.~\ref{fig:ent2b3}).
The maximal entanglement is not the sum of the maximal qutrit
entanglement and the maximal orbital entanglement because the orbital
motion leads to particle number fluctuations and reduces the spin entanglement (see
Fig.~\ref{fig:ent2b5}). The maximal orbital entanglement is realized in
the limit of strong tunneling; the maximal spin entanglement
corresponds to the maximally localized state, i.e., $|\psi_2 \rangle$.

\subsection{Three spin-1 bosons} 

\begin{figure}[htb]
 \begin{center}
\includegraphics[width=0.4\textwidth]{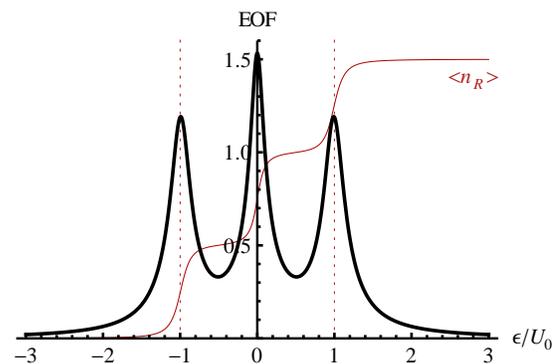}
\caption{(Color online) EOF between two wells for three bosons with
  antiferromagnetic interactions ($t/U_0=0.1$ and $U_2/U_0=0.1$) for
  the total spin $S_{tot}=1$. \label{fig:ent3b1}}
 \end{center}
 \end{figure}

\begin{figure}[ht]
 \begin{center}
 \includegraphics[width=0.4\textwidth]{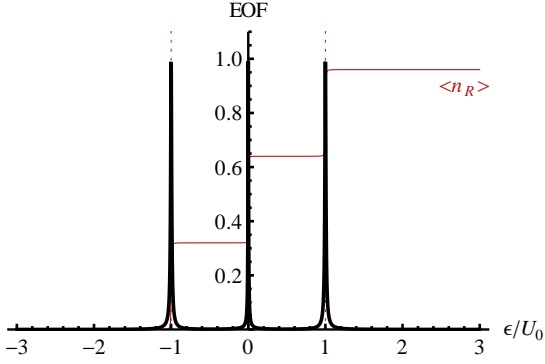}
\caption{(Color online) EOF between two wells for three bosons with
  antiferromagnetic interactions ($t/U_0=0.005$ and $U_2/U_0=0.1$) for
  the total spin $S_{tot}=1$. \label{fig:ent3b2} }
 \end{center}
 \end{figure}

The EOF between the two sites is presented in Fig.~\ref{fig:ent3b1}
for $t/U_0=0.1$ and in Fig.~\ref{fig:ent3b2} for very weak
tunneling. In contrast to the case of two bosons, in the
weak-tunneling case the system is not entangled for large intervals of
the energy offset $\epsilon$.


To quantify this effect, we analyze the EOF again in detail. The spins
$\vec S_L$ and $\vec S_R$ couple to a total spin, for which three absolute 
values are possible, $S_{tot} \in \{ 1,2,3\}$.

\begin{figure}[h]
 \begin{center}
 \includegraphics[width=0.4\textwidth]{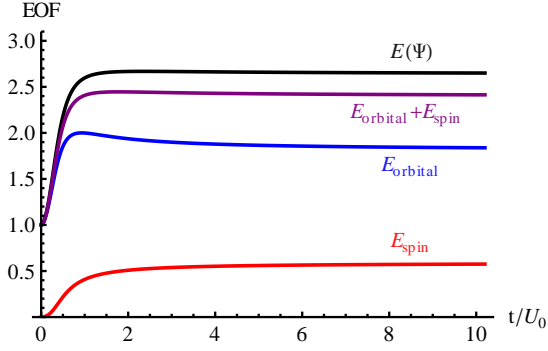}
\caption{(Color online) EOF, $E_{\text{spin}}$, $E_{\text{orbital}}$ and
  $E_{\text{spin}}+E_{\text{orbital}}$ between two symmetric wells for three bosons with
  antiferromagnetic interactions ($U_2/U_0 = 0.1$) in a symmetric
  double-well potential ($\varepsilon/U_0=0$). \label{fig:ent3b5} }
 \end{center}
 \end{figure}

It is obvious that Eq.~(\ref{2bedecompose}) is applicable for
$S_{tot}=2$ and $S_{tot}=3$. The interesting case is $S_{tot}=1$. Each
state with quantum number $S_{tot}=1$ can be written as
\begin{eqnarray}
\label{3b}
| \Psi\rangle = \sum_i^6 c_i | \psi_i\rangle.
\end{eqnarray}
Only two of the basis vectors contain true spin entanglement,

\begin{eqnarray*}
| \psi_2\rangle =|((2,1),&(2,1)&,1)\rangle= \alpha |2,2,0;1,1,0\rangle \\
&+& \beta |2,2,1;1,1,-1\rangle + \gamma |2,2,-1;1,1,1\rangle \\
| \psi_4\rangle=|((1,2),&(1,2)&,1)\rangle= \alpha |1,1,0;2,2,0\rangle \\
&+& \beta |1,1,1;2,2,-1\rangle + \gamma  |1,1,-1;2,2,1\rangle 
\end{eqnarray*}
with $\alpha=-\sqrt{\frac{2}{5}}$ and $\beta=\gamma=
\sqrt{\frac{3}{10}}$. Any superposition of $| \psi_2\rangle$ and $|
\psi_3\rangle$ can be written as
\begin{eqnarray}
\label{psi2psi3}
c_2 &| \psi_2\rangle &+ c_3 | \psi_3\rangle    \nonumber \\
&= &\sqrt{c_2^2\alpha^2+c_3^2} |L\rangle\otimes |1,1,0\rangle_R   \nonumber \\
&+& c_2\beta |2,2,1;1,1,-1\rangle + c_2\gamma |2,2,-1;1,1,1\rangle
\end{eqnarray}
where $|L\rangle$ is the normalized function
$1/\sqrt{c_2^2\alpha^2+c_3^2}\ (c_2\alpha |2,2,0\rangle_L+c_3
|2,0,0\rangle_L)$, which is orthogonal to the other vectors appearing
in Eqs.~(\ref{psi2psi3}) and (\ref{3b}). Therefore, the decomposition
Eq.~(\ref{psi2psi3}) is a Schmidt decomposition and the full
entanglement of formation of $|\Psi\rangle$ can be calculated,
\begin{eqnarray}
\label{3be}
E(|\Psi \rangle) &=& - c_1^2 \log_2 c_1^2 - 
\left( c_2^2\alpha^2+c_3^2\right)\log_2 \left( c_2^2\alpha^2+c_3^2\right)  \nonumber\\
&-& \left( c_2^2\beta^2\right)\log_2 \left( c_2^2\beta^2\right) - \left( c_2^2\gamma^2\right)\log_2 \left( c_2^2\gamma^2\right)  \nonumber\\
 &-&\left( c_4^2\alpha^2+c_5^2\right)\log_2 \left( c_4^2\alpha^2+c_5^2\right) - \left( c_4^2\beta^2\right)\log_2 \left( c_4^2\beta^2\right) \nonumber\\ 
 &-& \left( c_4^2\gamma^2\right)\log_2 \left( c_4^2\gamma^2\right) -
c_6^2 \log_2 c_6^2\: .
\end{eqnarray}
It is possible to decompose the entanglement into different contributions and to
generalize the expressions for $E_{\text{orbital}}$ and $E_{\text{spin}}$ in
Eq.~(\ref{2bedecompose}). To calculate the orbital entanglement we
construct the orbital wave function and use this to get the
EOF of the reduced density matrix.  The orbital wave
function is
\begin{eqnarray*}
&& |\Psi\rangle_{\text{orbital}} = \\
&&c_1 |3,0\rangle+ \sqrt{c_2^2 + c_3^2 }|2,1\rangle+ \sqrt{c_4^2 + c_5^2 }|1,2\rangle+c_6 |0,3\rangle,
\end{eqnarray*}
where the quantum numbers refer to $|n_L,n_R\rangle$. So the orbital
entanglement of formation between the left and the right well is given
by
\begin{eqnarray}
\label{3beorbital}
E_{\text{orbital}}=&-&c_1^2 \log_2 c_1^2 - (c_2^2 +c_3^2)\log_2 (c_2^2 +c_3^2)  \nonumber \\
&-& (c_4^2 +c_5^2)\log_2 (c_4^2 +c_5^2)- c_6^2 \log_2 c_6^2 
\end{eqnarray}
The spin wave function is given by
\begin{eqnarray}
\label{3bspinpsi}
|\Psi\rangle_{\text{spin}} = \sqrt{c_1^2 + c_5^2 } |\{1,0\},1\rangle+ c_2 |\{2,1\},1\rangle  \nonumber  \\
+ \sqrt{c_3^2 + c_6^2 }|\{0,1\},1\rangle+c_4 |\{1,2\},1\rangle,
\end{eqnarray}
where the quantum numbers refer to $|\{S_L,S_R\},S_{tot}\rangle$. The
EOF of these orthonormal basis vectors is
$E(|\{1,0\},1\rangle)=E(|\{0,1\},1\rangle)=0$ and
$E(|\{2,1\},1\rangle)=E(|\{1,2\},1\rangle)=-\alpha^2 \log_2
\alpha^2-\beta^2 \log_2 \beta^2-\gamma^2 \log_2 \gamma^2$. Therefore, the
EOF due to spin entanglement is
\begin{eqnarray}
\label{3bespin}
E_{\text{spin}}= c_2^2 E(|\psi_2\rangle) + c_4^2 E(|\psi_4\rangle). 
\end{eqnarray}
Note that
\begin{eqnarray}
\label{3babschaetz}
 & &\left( c_2^2\alpha^2+c_3^2\right)\log_2 \left( c_2^2\alpha^2+c_3^2\right)+ \left( c_2^2\beta^2\right)\log_2 \left( c_2^2\beta^2\right)  \nonumber \\
& &+ \left( c_2^2\gamma^2\right)\log_2 \left( c_2^2\gamma^2\right)    \nonumber \\
& \leq& c_3^2\log_2 \left( c_2^2+c_3^2\right)+\left(c_2^2\right) \log_2 \left(1+\frac{c_3^2}{ c_2^2}\right)   \nonumber \\
& +&c_2^2 \log_2 c_2^2 -  c_2^2  E(|\psi_2\rangle)   \nonumber  \\
&=& \left(  c_2^2 + c_3^2\right)  \log_2 \left( c_2^2+c_3^2\right) -  c_2^2  E(|\psi_2\rangle),
\end{eqnarray}
where $\log(1+d z) \leq d \log(1+z)$ for $d\geq 1$ and $z\geq 0$ has
been used.  Because of Eq.~(\ref{3babschaetz}) the entanglement of
formation is bounded from below (see Fig.~\ref{fig:ent3b5}),
\begin{eqnarray}
\label{3bedecompose}
 E(|\Psi \rangle) \geq E_{\text{orbital}} + E_{\text{spin}}
\end{eqnarray}

\subsection{Arbitrary number of bosons}

Let $\Psi$ be a wave function that describes the state of $N$
bosons. This wave function can be written in terms of a basis, which
is ordered according to the occupation numbers $N_L$ and
$N_R$, the spin in the left well $S_L$ and the spin in the right well $S_R$, and
the total spin $S_{tot}$;
\begin{eqnarray}
\label{Nbbasis}
|\Psi\rangle=\sum_{n=1}^D c_n |\phi_n \rangle,
\end{eqnarray}
where $\sum_n c_n^2 =1 $ and $D \geq N$ is the dimension of the basis.
We can rearrange this sum by sorting it according to the occupation numbers,
\begin{eqnarray*}
|\Psi\rangle=\sum_{m=0}^N |\psi_m \rangle,
\end{eqnarray*}
where $ |\psi_m \rangle$ is the part of the wave function belonging to
$N_L=m$. If $N(m)$ is the number of basis vectors belonging to
$N_L=m$, $ |\psi_m \rangle$ is given by
\begin{eqnarray}
\label{Nm}
| \psi_m \rangle = \sum_{i=1}^{N(m)} c(m)_i | \phi(m)_i \rangle,
\end{eqnarray}
where $\sum_{i=1}^{N(m)} c(m)_i^2 \leq 1$ and $c(m)_i$ denote the
coefficients $c_i$ that belong to $N_L=m$.  Now it is possible to
generalize Eqs.~(\ref{3beorbital}) and (\ref{3bespin}) and to
define the orbital EOF,
 \begin{eqnarray}
 \label{Neorbital}
  E_{\text{orbital}}=-\sum_{m=0}^N \sum_{i=1}^{N(m)} c(m)_i^2  \log_2 
\sum_{i=1}^{N(m)} c(m)_i^2  
  \end{eqnarray} 
 and the spin EOF
 \begin{eqnarray}
 \label{espin}
 E_{\text{spin}}=\sum_{n=1}^D c_n^2 E(|\phi_n\rangle).
 \end{eqnarray}
It is not necessary to to specify which basis vectors in
Eq.~(\ref{Nbbasis}) belong to which angular momentum configuration, such as in Eq.~(\ref{3bspinpsi}) because the total spin-entanglement
entropy can be written as a sum over all basis vectors.
 
In this section we prove that Eq.~(\ref{3bedecompose}) is true for any
number of bosons in a double well:
\begin{eqnarray}
\label{Nbedecompose}
 E(|\Psi \rangle) \geq E_{\text{orbital}} + E_{\text{spin}}\:.
\end{eqnarray}
$E(|\Psi \rangle) $ decomposes in a sum over $m$: $E(|\Psi
\rangle)=\sum_m^N E(|\psi_m \rangle) $. It is possible to write down
the EOF for each $|\psi_m\rangle$ in the following
way:
\begin{eqnarray}
\label{Nmbedecompose}
&&E(|\psi_m \rangle)=  \\
 &-&\sum_j \left( \sum_i \alpha(m)_{ij}^2 c(m)_i^2\right) \log_2  
\left( \sum_i \alpha(m)_{ij}^2 c(m)_i^2\right),\nonumber  
\end{eqnarray}
which defines a basis for each vector $\phi(m)_i$,
\begin{eqnarray*}
|\phi(m)_i \rangle &=&\sum_k \sum_l a(m)_{i k} a(m)_{i l} | L_k 
\rangle \otimes |R_l\rangle\\
&=&\sum_j \alpha(m)_{ij} |L,R\rangle_j\:,
\end{eqnarray*}
where $\sum_j \alpha_{ij}^2=1$. To prove Eq.~(\ref{Nbedecompose}) for
any number of bosons, it is necessary and sufficient to show that
Eq.~(\ref{Nbedecompose}) is true for each $E(|\psi_m \rangle)$, i.e.,
\begin{eqnarray}
\label{ersteUn}
&&  \sum_j \left( \sum_i \alpha_{ij}^2 c_i^2\right) \log_2  
\left( \sum_i \alpha_{i j}^2 c_i^2\right)   \\
  \leq & & 
  \left( \sum_i c_i^2 \right) \log_2 \left( \sum_i c_i^2 \right) 
+ \sum_i c_i^2 \sum_j \alpha_{ij}^2 \log_2 \alpha_{ij}^2\:.\nonumber 
\end{eqnarray}
The term on the left-hand side can be rearranged,
\begin{eqnarray}
  \sum_j &&\left( \sum_i \alpha_{ij}^2 c_i^2\right) \log_2  
\left( \sum_i \alpha_{ij}^2 c_i^2\right)  \nonumber \\
 &=&
  \sum_j \sum_i \left(\alpha_{ij}^2 c_i^2\right) \log_2  
\left(\alpha_{ij}^2 c_i^2\right)  \nonumber  \\ 
  &+& \sum_j \sum_i  \left(\alpha_{ij}^2 c_i^2\right) \log_2  
\left(\frac{\sum_n \alpha_{nj}^2  c_n^2 }{\alpha_{ij}^2 c_i^2}\right) \:,
\end{eqnarray}
as well as the term on the right-hand side,
\begin{eqnarray*}
&&\left( \sum_i c_i^2 \right) \log_2 \left( \sum_i c_i^2 \right) + \sum_i c_i^2 \sum_j \alpha_{ij}^2  \log_2\alpha_{ij}^2  \\ 
 &=&  \left( \sum_i c_i^2 \right) \log_2 \left( \sum_i c_i^2 \right) - \sum_i c_i^2 \log_2 c_i^2   \\
  &+& \sum_i \sum_j \left(\alpha_{ij}^2 c_i^2 \right) \log_2 \left( \alpha_{ij}^2  c_i^2 \right)\:.
\end{eqnarray*}
Note that, due to Jensen's inequality,
\begin{eqnarray*}
 \sum_j  \left(\alpha_{ij}^2 c_i^2\right) \log_2   \left(\frac{\sum_n \alpha_{nj}^2  c_n^2 }{\alpha_{ij}^2 c_i^2}\right)\\
  \leq   c_i^2 \log_2   \left(\frac{ \sum_j  \sum_n \alpha_{nj}^2  c_n^2 }{c_i^2}\right)   \\
  =   c_i^2 \log_2  \left( \sum_{n}  c_n^2 \right) -  c_i^2 \log_2  c_i^2,
\end{eqnarray*}
Eq.~(\ref{ersteUn}) is fulfilled and, therefore, Eq.~(\ref{Nbedecompose}).

\subsection{Comparison with the entanglement of particles}

The amount of entanglement shared between two parties might be lowered
by superselection rules~\cite{bartlett03}. In case two parties share
$N$  particles and a particle superselection rule
applies, the extractable bipartite entanglement, i.e., the degree of
entanglement one can entangle two initially not entangled quantum
registers located at $A$ and $B$, is given by the entanglement of
particles~\cite{wiseman03},
\begin{eqnarray*}
E_P\left( | \Psi_{AB}\rangle\right) \equiv \sum_n P_n \ E
\left(| \Psi_{AB}^{(n)}\rangle\right),
\end{eqnarray*}
where $|\Psi_{AB}^{(n)}\rangle$ is $|\Psi_{AB}\rangle$ projected
onto the subspace of fixed local particle number, i.e., $n$ particles
for one party and $n-1$ for the other.

The entanglement of particles for two bosons in a double well is given
by $E_{\text{spin}}$ in Eq.~(\ref{2bedecompose}). For three bosons the case
$S_{tot}=2$ and $S_{tot}=3$ is trivial, but the case $S_{tot}=1$ is
more interesting and is examined. To calculate $E_P$, we write
down the projection obeying local particle superselection rules. The
projections onto $n_L=3$ and $n_L=0$ are trivial and do not contribute
to $E_P$. The projection onto $n_L=2$ leads to Eq.~(\ref{psi2psi3})
with $P_2=c_2^2+c_3^2$. The entanglement contained in this state is
given by
\begin{eqnarray*}
E\left( | \Psi_{LR}^{(2)} \rangle \right)&=&- \frac{ c_2^2\alpha^2+c_3^2}{ c_2^2+c_3^2} \log_2 \frac{ c_2^2\alpha^2+c_3^2}{ c_2^2+c_3^2} \\
&-& \frac{ c_2^2\beta^2}{ c_2^2+c_3^2} \log_2 \frac{ c_2^2\beta^2}{ c_2^2+c_3^2}-
\frac{ c_2^2\gamma^2}{ c_2^2+c_3^2} \log_2 \frac{ c_2^2\gamma^2}{ c_2^2+c_3^2}
\end{eqnarray*}
 and thereby contributes 
\begin{eqnarray*}
P_2 \ E | \Psi_{LR}^{(2)} \rangle &=&- 
\left( c_2^2\alpha^2+c_3^2 \right) \log_2\left( c_2^2\alpha^2+c_3^2 \right) \\
&-& \left( c_2^2\beta^2\right)\log_2 \left( c_2^2\beta^2\right)- 
\left( c_2^2\gamma^2\right)\log_2 \left( c_2^2\gamma^2\right)\\
&+&\left(c_2^2+c_3^2 \right) \log_2 \left(c_2^2+c_3^2 \right)
\end{eqnarray*}
to $E_P$.
A comparison with Eq.~(\ref{3be}) shows that the equation,
\begin{eqnarray}
\label{epun}
E(|\Psi \rangle) = E_{\text{orbital}} + E_P
\end{eqnarray}
holds for three bosons. This equation also is true for higher boson
numbers. The contribution of the state (\ref{Nm}) to $E_P$ is given by
\begin{eqnarray*}
&&P_m \ E | \Psi_{LR}^{(m)} \rangle =\\
&-&\sum_j \left( \sum_i \alpha(m)_{ij}^2 c(m)_i^2\right) \log_2  
\left( \sum_i \alpha(m)_{ij}^2 c(m)_i^2\right) \\
&+& \left( \sum_i c(m)_i^2\right) \log_2  \left( \sum_i c(m)_i^2\right)
\end{eqnarray*}
A comparison with Eq.~(\ref{Neorbital}) shows that Eq.~(\ref{epun}) indeed
holds for all boson numbers.

The necessity to take a superselection rule into account may arise due
to several reasons. In some cases the phase between states with
different local particle occupation numbers is not well
defined~\cite{dowling06}. Consider the bipartite state
\begin{eqnarray}
|\psi_{\theta}\rangle_{AB} =\sqrt{\frac{1}{2}}
\left(| 1,0 \rangle + e^{i \phi} | 0,1 \rangle \right).
\end{eqnarray} 
In case there is no shared reference frame and no tunneling 
between the two parties the phase is not accessible experimentally and
the state is indistinguishable from an incoherent mixture,
\begin{eqnarray}
\rho_{AB}=\frac 1 2 \left( | 1,0\rangle \langle 1,0  | +  
| 0,1\rangle \langle 0,1  | \right)\:.
\end{eqnarray}
Whenever one is concerned with the occupation number of massive
particles, the detailed properties of the system determine which local operations and classical communication (LOCCs) are allowed: If
tunneling is forbidden, LOCCs will conserve the local particle
number. In this case a local particle number superselection rule must
be taken into account.  A more trivial example is the case of a
superselection rule for the total particle number~\cite{cramer11}.

In our model~(\ref{mitspin}) the phase is well defined due to the
finite-tunneling amplitude.  The amount of orbital entanglement
$E_{\text{orbital}}$ depends directly on the particle fluctuations
caused by the tunneling between the sites.  In the absence of
tunneling, the orbital entanglement vanishes and the superselection
rule for the local particle number is effectively enforced.

\subsection{Creation of entanglement structures}

In the case of two spin-1 bosons in a double well the state of total spin
zero ($S_{tot}=|\vec S_L + \vec S_R |=0$) is singled out. First, it can
be separated from the $S_{tot}=2$ state due to a different particle
distribution within the double well in the vicinity of the
single-particle tunneling resonance (i.e. $\epsilon / U_0=
0.5$). Second, it represents the two-qutrit singlet state and thereby
contains the maximal qutrit entanglement of $\log_2 3$. This
distinguishes the qutrit entanglement from the qubit entanglement where
the singlet state and the triplet $(S_{tot})_z=0$ state contain the
same amount of entanglement.

This can be used to create specific entanglement structures in 2D
optical superlattices (see Fig.~\ref{fig:two}).

\section{Conclusion}

We have analyzed the two-site Bose-Hubbard model for spin-1 atoms
explicitly for small numbers of bosons. Starting from the explicit
form of the Hamiltonian, we have discussed the physics of the bosonic
staircases. We also have studied the effect of magnetic fields.  In
the following we have examined the bipartite entanglement for the
two-site Bose-Hubbard model. We have analyzed the contribution of
orbital and spin degrees of freedom and have derived a lower bound of the
total entanglement, which is the sum of the orbital entanglement and
the spin entanglement.  We compared the entanglement of particles and
thereby elucidated the meaning of orbital entanglement and of
superselection rules for the local particle number.

The staircases for different total spins establish a
correspondence of the spatial motion and the spin
configuration. Because the $S_{tot}=0$ singlet state of two bosons
does contain more entanglement than the other eigenstates of the
system this correspondence can be used to construct an entanglement
witness in the system: In case one detects the typical spatial
behavior of the $S_{tot}=0$ state, one can conclude to have its
entanglement.  With the help of fluorescence imaging, it is also
possible to depopulate doubly occupied sites in the lattice and
thereby to build a spin filter.

We have discussed entanglement between the sites, not the entanglement
between the individual atoms. Even for an occupancy of one, i.e. one
atom per site, these are different quantities, because the bosons are
indistinguishable. Recently it was proposed to measure the
entanglement between (spinless) bosons in an optical
lattice~\cite{cramer11} by standard time-of-flight measurements. Such
measurements do not preserve the information about the entanglement
between the individual sites. There are other possibilities for examining
these systems experimentally.  First, it is possible to estimate the
entanglement by measurements of the atom positions because these
correspond to specific spin configurations as we have
demonstrated. These atom positions can be determined by standard
time-of-flight measurements or direct fluorescence detection of
individual sites~\cite{sherson10}. Furthermore, it is possible to
detect the spin configurations directly in a nondemolishing way with
the help of the quantum Faraday effect~\cite{eckert08}.  Furthermore,
it may be possible to relate the entanglement to additional observable
experimental quantities, such as magnetization fluctuations in one of
the wells, in analogy to what has been discussed for noninteracting
particles~\cite{song11}. We plan to explore this question in future
papers.

\section{Acknowledgments}

We would like to thank R. Fazio and T.L. Schmidt for discussions. 
This paper was financially supported by
the Army Research Office with funding from the DARPA
OLE program, Harvard-MIT CUA, NSF Grant 
No. DMR-07-05472, AFOSR Quantum Simulation MURI, AFOSR MURI
on Ultracold Molecules, and the ARO-MURI on Atomtronics, 
and by the the Swiss SNF, the NCCR Nanoscience, and the NCCR
Quantum Science and Technology.

\appendix

\section{Coupling of $n$ spin-1 bosons}
\label{appendix:coupling}

The coupling of two spin-1 atoms is given in standard textbooks. The
coupling of $n$ spin-1 atoms to a total spin $\vec S$ with a
$z$-projection $S_z$ in terms of $n$ single spins $ S_{iz}$ is
calculated by the diagonalization of $\vec S^2$.

For three spin-1 atoms we give the connection of the basis vectors
ordered according to $ S_{iz}$ and the basis vectors ordered according
to $S$ and $S_z$,
\begin{eqnarray*}
|S=3,\ S_z=3 \rangle &=&\left| 0_{-1},0_0,3_1\right\rangle ,\\
|S=3,\ S_z=2 \rangle &=& \left|0_{-1},1_0,2_{1}\right\rangle,\\
|S=3,\ S_z=1 \rangle &=& (2 \left|0_{-1},2_0,1_1\right\rangle +\left|1_{-1},0_{0},2_{1}\right\rangle)/\sqrt{5},\\
|S=3,\ S_z=0 \rangle &=&\sqrt{\frac{2}{5}} \left|0_{-1},3_0,0_1\right\rangle +\sqrt{\frac{3}{5}} \left|1_{-1},1_{0},1_{1}\right\rangle ,\\
|S=3,\ S_z=-1 \rangle &=&(2 \left|1_{-1},2_{0},0_{1}\right\rangle +\left|2_{-1},0_{0},1_{\hat{1}}\right\rangle)/\sqrt{5},\\
|S=3,\ S_z=-2 \rangle &=&\left|2_{-1},1_{0},0_{1}\right\rangle ,\\
|S=3,\ S_z=-3 \rangle &=&\left|3_{-1},0_0,0_1\right\rangle ,\\
|S=1,\ S_z=1 \rangle &=&(-\left|0_{-1},2_{0},1_{1}\right\rangle +2 \left|1_{-1},0_0,2_1 \right\rangle)/\sqrt{5},\\
|S=1,\ S_z=0 \rangle &=&-\sqrt{\frac{3}{5}} \left|0_{-1},3_0,0_1\right\rangle +\sqrt{\frac{2}{5}} \left|1_{-1},1_{0},1_{1}\right\rangle ,\\
|S=1,\ S_z=-1 \rangle &=& (-\left|1_{-1},2_0,0_1\right\rangle +2 \left|2_{-1},0_0,1_1 \right\rangle)/\sqrt{5}
\end{eqnarray*}


%

\end{document}